\documentclass[10pt, conference]{IEEEtran}
\IEEEoverridecommandlockouts
\usepackage{cite}
\usepackage{amsmath,amssymb,amsfonts}
\usepackage{algorithmic}
\usepackage{textcomp, multirow}
\usepackage{color}

\def\BibTeX{{\rm B\kern-.05em{\sc i\kern-.025em b}\kern-.08em
    T\kern-.1667em\lower.7ex\hbox{E}\kern-.125emX}}

\ifCLASSINFOpdf
  \usepackage[pdftex]{graphicx}
  \graphicspath{{./}}
  \DeclareGraphicsExtensions{.pdf,.jpg,.png,.xps}
\else
\fi

\IEEEoverridecommandlockouts
\IEEEpubid{\makebox[\columnwidth]{978-1-5090-0223-8/16/\$31.00~\copyright~2016 European Union \hfill} \hspace{\columnsep}\makebox[\columnwidth]{ }}
\begin{document}

\title{Balancing the Migration of Virtual Network Functions with Replications in Data Centers\\
}

\author{\IEEEauthorblockN{Francisco Carpio, Admela Jukan and Rastin Pries}
	\IEEEauthorblockA{Technische Universit{\"a}t Braunschweig, Germany, Nokia Bell Labs, Germany}
	\IEEEauthorblockA{Email:\{f.carpio, a.jukan\}@tu-bs.de, rastin.pries@nokia-bell-labs.com}
}

\maketitle

\begin{abstract}
The Network Function Virtualization (NFV) paradigm is enabling flexibility, programmability and implementation of traditional network functions into generic hardware, in form of the so-called Virtual Network Functions (VNFs). Today, cloud service providers use Virtual Machines (VMs) for the instantiation of VNFs in the data center (DC) networks. To instantiate multiple VNFs in a typical scenario of Service Function Chains (SFCs), many important objectives need to be met simultaneously, such as server load balancing, energy efficiency and service execution time. The well-known \emph{VNF placement} problem requires solutions that often consider \emph{migration} of virtual machines (VMs) to meet this objectives. Ongoing efforts, for instance, are making a strong case for migrations to minimize energy consumption, while showing that attention needs to be paid to the Quality of Service (QoS) due to service interruptions caused by migrations. To balance the server allocation strategies and QoS, we propose using \emph{replications} of VNFs to reduce migrations in DC networks. We propose a Linear Programming (LP) model to study a trade-off between replications, which while beneficial to QoS require additional server resources, and migrations, which while beneficial to server load management can adversely impact the QoS. The results show that, for a given objective, the replications can reduce the number of migrations and can also enable a better server and data center network load balancing.
\end{abstract}


\section{Introduction}
Recently, Network Function Virtualization (NFV) has emerged as a new paradigm that virtualizes the traditional network functions and places them into generic hardware inside the network or in data centers, as opposed to the traditional designated hardware. Since a single Virtual Network Functions (VNF) cannot provide full services, multiple VNFs are commonly linked in a sequence order, known as Service Function Chains (SFCs), and placed into the network which introduces the so-called VNF placement problem. The placement of the VNFs can happen either in data centers (DC) or by deploying single servers or clusters of servers inside the network. Cloud service providers today are increasingly using such VNF placements that can enable flexible usage of computing and storage in their data centers by placing VNF instances (VNFIs) into Virtual Machines (VMs) or containers  that can migrate depending on the requirements. During the high service demands, the VM or container migrations can be used with the objective to balance the server and network loads, while assuring an acceptable Quality of Service (QoS). 
 
\begin{figure}[!t]
	\centering
	\includegraphics[width=0.5\textwidth]{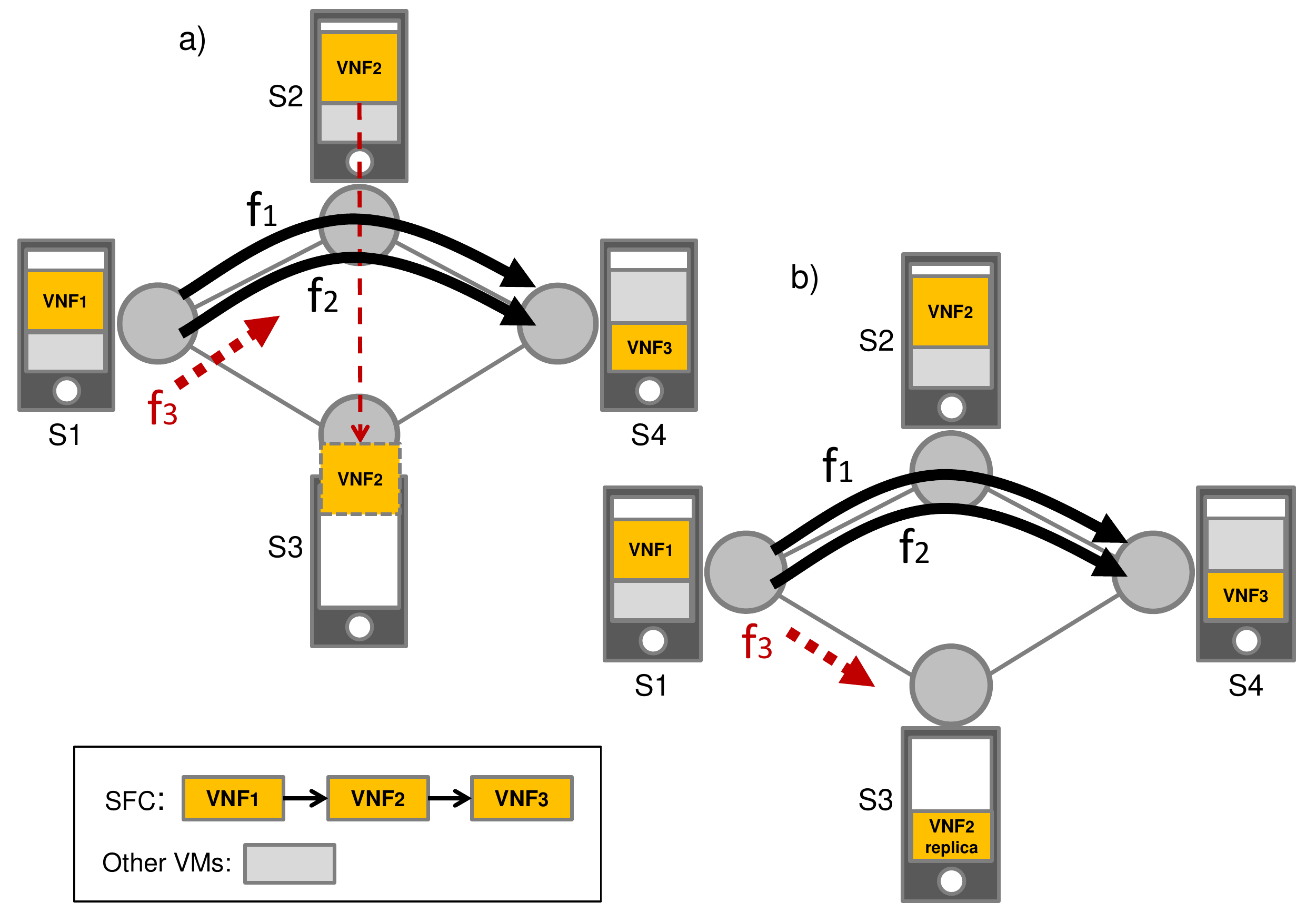}
	\caption{A comparison between migration and replication.}
	\label{case}
\end{figure}

The main factor impacting the QoS is, in general, the migration of \emph{stateful VNFs} that has to deal with the problem of transferring the internal flow states, whereby additional mechanisms and protocols are necessary to keep the states during the migration process, which makes it non-trivial. Current research proposes the migration of the entire VM (and not individual VNFs), which is a feasible and reliable technology used in modern DCs. However, it introduces additional challenges since a single VM can contain multiple VNFs and not all VNFs can be migrated at the same time. This limitation on flexibility can be even harder when multiple VNFs are concatenated in an SFC. To address this problem in practical terms, current solutions consider either one VNF per VM only, and use VM live migration for the migration of VNFs between servers, or one VNF per container. While these solutions are valid, the migration of VNFs still impacts the QoS perceived by the end-user due to the stop/start operations. Therefore, reducing the migrations of VNFs generally reduces the negative impacts on live services and is a desirable goal. 

To address this challenge, we propose using \emph{replications} to reduce the VNF migrations to achieve the right balance between the network and server utilization, thus minimizing the service interruptions. We illustrate the idea in Fig. \ref{case}. Let us assume one VNF per VM, with the traffic demands as network flows $f_1$ and $f_2$ in an SFC containing VNF1-VNF2-VNF3. As shown in Fig. \ref{case}a, the network traffic traverses the servers S1, S2 and S4, respectively, and the corresponding links. Assume now a third flow is required for the same service in VNF2, leading to an overload of S2 due to an increased size of VNF2. To address this, we can migrate the VNF2 to an underused server S3, and all three traffic flows migrate to S3 as well. While this solution is valid, the migration of VNF2 affects the QoS perceived by flows $f_1$ and $f_2$ due to the stop/copy operations. Fig. \ref{case}b illustrates the same scenario with replications. Here, before flow $f_3$ demand, the VNF2 is replicated into S3, by creating a new VNF2 (replica). Then, this replica provides service to flow $f_3$. The initial two flows remain unaffected, and we not only load balance the servers but also the network by increasing the number of admissible paths able to provide service. 

In this paper, we propose a VNF replication method and study its usefulness in two different types of DCNs: 1) the traditional fat-tree topology using the default ECMP forwarding protocol, and 2) a more recent leaf-spine topology, where the SDN technology is used. Two related but different Linear Programming (LP) models have been developed and applied to both case studies and for both cases of the migration and replication. The objective, for both methods, is to load balance the server and network link utilization. We analyze and compare the two methods in terms of number of replicas, which requires additional resources, and migrations, which create QoS issues. While quantifying the overhead introduced by the creation of new VMs as replicas, in this paper, we show that, for a given objective, replications can reduce the number of required migrations, while at the same time, load balance the server and link utilization.

The rest of the paper is organized as follows. Section II presents related work. Section III describes the background on migration and replication in data centers.  Section IV formulates the analytical optimization model. Section V analyzes the performance and Section VI concludes the paper.

\section{Related Work and Our Contribution}

VNF migration has been addressed previously in various contexts. A typical example are migrations during the off-peak hours in DCNs to reduce energy consumption. As presented in \cite{Eramo2017a}, the goal to migrate the VNFs to a few servers only, such that the remaining servers can be switched off. To address the issue of service interruptions that negatively impact QoS during migrations, \cite{Eramo2017} derived a trade-off between the power consumption and QoS degradation to determine whether a migration is appropriate. VM migration is also necessary during high demand in DCN to meeting the increased requirements for computing jobs and the related network resources. As discussed in \cite{Xia2016a}, further research in VNF migration is needed, not only related to the VNF migration protocols and methods as in \cite{Xia2016}, or the reduction of  migration time such as in \cite{Zhang2017} \cite{Cho2017}, but also on optimization of further parameters. For instance, while optimizing migrations is important, migration requires additional methods and systems for server load balancing as well as balancing of network resources, which is an important factor for DCN infrastructure planning and operation.  To this end, past work has been proposed to solving the so-called \emph{VNF placement} problem, by allocating VNFs to load balance the network or server utilization, or both \cite{Carpio2016a, Ghaznavi2015, Carpio2017}. At the same time, however, these work do not consider migration. 

In this paper, our novel contribution is to balance migration with replications, with consideration of server and network load balancing. We propose a novel analytical model to balance the number of migrations with replications, and thus, minimize the number of migration and minimize the impact on the QoS associated with migrations, while at the same time making sure that replication is applied only when a comparably large benefit can be achieved from the server and network balancing perspective. Another novel contribution is the consideration of service chains for migration and replication, which to the best of our knowledge no past work has addressed. The latter observation is important, since although VM and VNF migration can be seen as a similar problem, -- whereby VM migration has been studied intensely, there are some remarkable differences. As we pointed out, VNFs are service-chained, and migration of one VNF in a chain interrupts the entire service chain. Second, while a VM is considered a unique entity to migrate, in NFV, multiple VNFs can be instantiated on the same VM, and thus one migration can involve mutliple VNFs. Thus, a VM migration affects all VNFs and their respective service chains. Since all related work makes a strong assumption that there is one VNF instance per VM, or container, this paper also studies service chains without multiple VNFs per VM.
\begin{figure}[!t]
	\centering
	\includegraphics[width=0.5\textwidth]{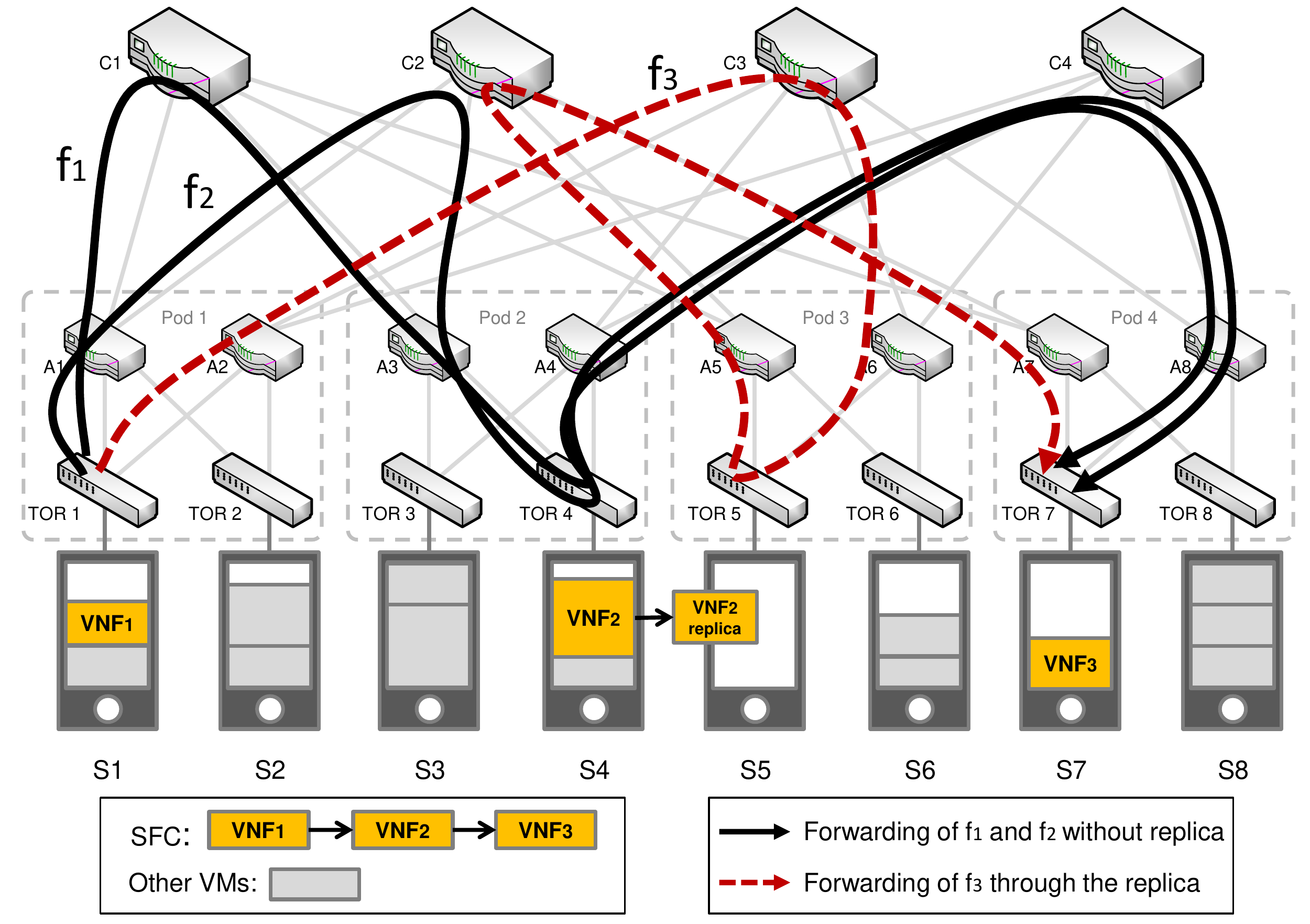}
	\caption{Forwarding of traffic demands in DCN scenarios}
	\label{rep}
\end{figure}

\begin{figure}[!t]
	\centering
	\includegraphics[width=0.49\textwidth]{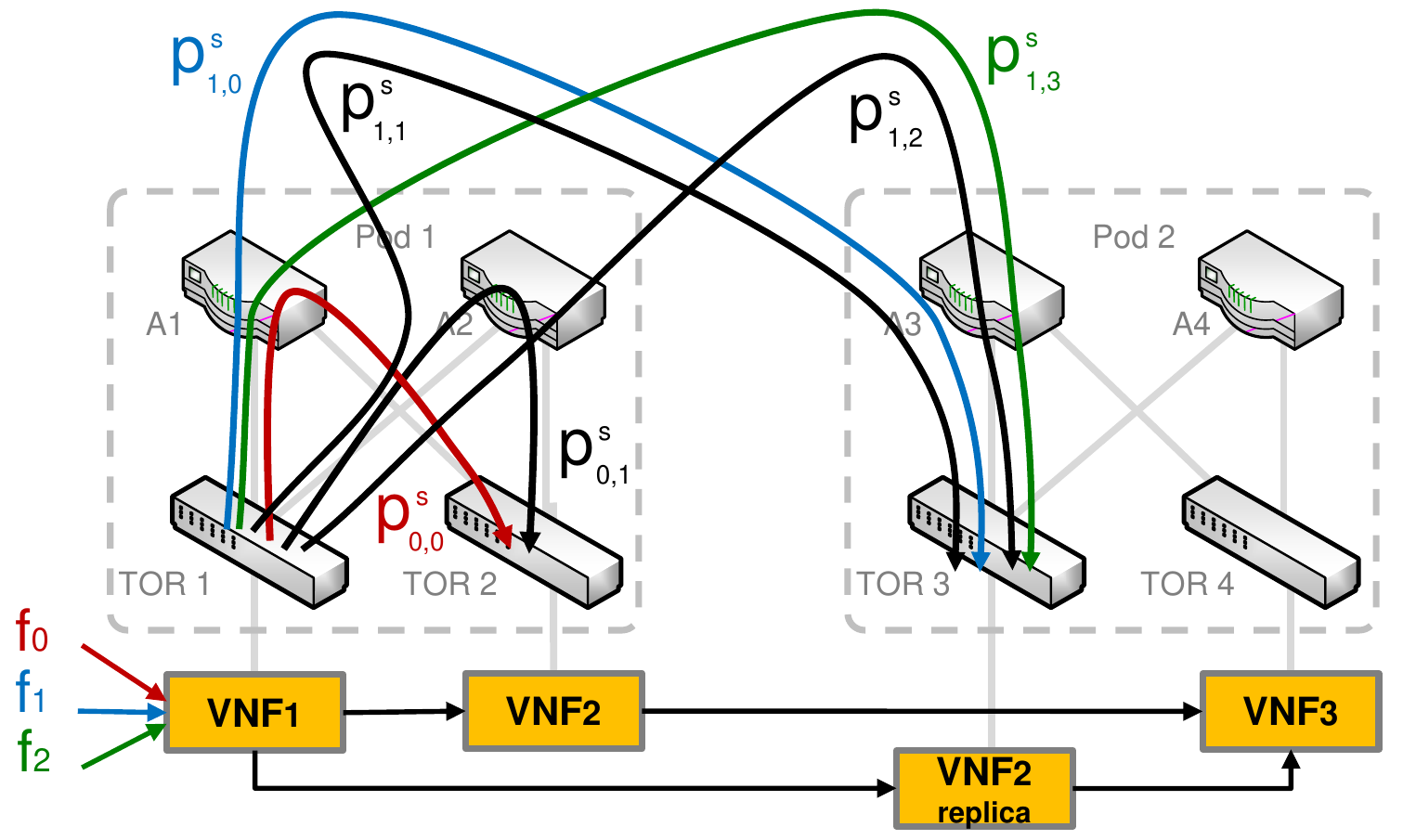}
	\caption{ECMP forwarding in fat-tree topology}
	\label{paths}
\end{figure}

\section{The Migration and Replication in Data Centers: ECMP and SDN} \label{problem}

This paper analyzes two different data center architectures: 1) the traditional fat-tree topology using the Equal Cost Multipath (ECMP) protocol, and 2) a recently proposed leaf-spine topology based on Software Defined Networking (SDN), as proposed by CORD \cite{cord}.

\subsection{The ECMP scenario}

The use of multipath ECMP forwarding protocol in traditional DCN scenarios does not present a major challenge from the perspective of network load balancing. However, when a server is overloaded and new traffic demands are generated for a service, the probability that different traffic demands can collide over the same path increases in fat-tree topologies. This scenario is illustrated in Fig. \ref{rep}, where two flows $f_1$ and $f_2$ are requiring for service and VNF\textsubscript{1}, VNF\textsubscript{2} and VNF\textsubscript{3} are allocated in servers S1, S4 and S7, respectively. Let us assume that, due to an intensive job allocated to VNF\textsubscript{2}, server S4 becomes overloaded. To address this, one solution is to migrate VNF\textsubscript{2} (or other VNFs) running on S4 towards an underused server, in this case S5. This solution requires to stop the current service execution to maintain the state of the VNF, while the migration is performed. Another solution (which we propose to use) is to replicate VNF\textsubscript{2} into the server S5, thus balancing the load of incoming traffic demands between the original VNF and the replica. Then, as shown in Fig. \ref{rep}, a newly incoming flow $f_3$ can be routed towards S5. This solution is able to better load balance the network traffic as compared with the migration, due to the increment of available paths between two end points. However, it increases the utilization of resources due to the overhead incurred by the creation of new VMs for the replicas.

Let us assume the traditional fat-tree topology with a set of pre-computed ECMP paths for all possible VNF locations, as input parameters to find the solution for migrations and replications. To understand how the model selects ECMP paths, an example is illustrated in Fig. \ref{paths}. Here, three VNFs provide service to three flows ($f_0$, $f_1$ and $f_2$) from source TOR\textsubscript{1} (Top of Rack, TOR) to the destination TOR\textsubscript{4}. Before the optimization of the placement of VNFs and their replicas, we need to pre-compute all ECMP paths and randomly pre-select for each source-destination pair of servers. This step is a necessary constraint due to the ECMP, because with ECMP, it is not possible to optimize the paths chosen between two servers, as the path is chosen by a hash function of the 5-tuple header fields. Therefore, following the previous example, a random function pre-selects for $f_0$ one ECMP path from TOR\textsubscript{1} to TOR\textsubscript{2} (i.e. $p_{0,0}^s$), one from TOR\textsubscript{1} to TOR\textsubscript{3} (i.e. $p_{1,2}^s$), one TOR\textsubscript{1} to TOR\textsubscript{4} and so on, for all combinations of pair of servers. Then, an optimization method for VNF placement can be applied depending on the objective function and considering the pre-select paths as input parameters. If the model decides to place the VNFs as shown in the Fig. \ref{paths}, forwarding $f_0$ to the VNF\textsubscript{2}, and $f_1$ and $f_2$ towards the replica of VNF\textsubscript{2}, then, the chosen paths are $p_{0,0}^s$, $p_{1,0}^s$ and $p_{1,3}^s$, respectively. In case the model decides to forward the traffic demands in a different way, the used paths will change according to the random pre-selection, as previously mentioned.

\begin{figure}[!t]
	\centering
	\includegraphics[width=0.49\textwidth]{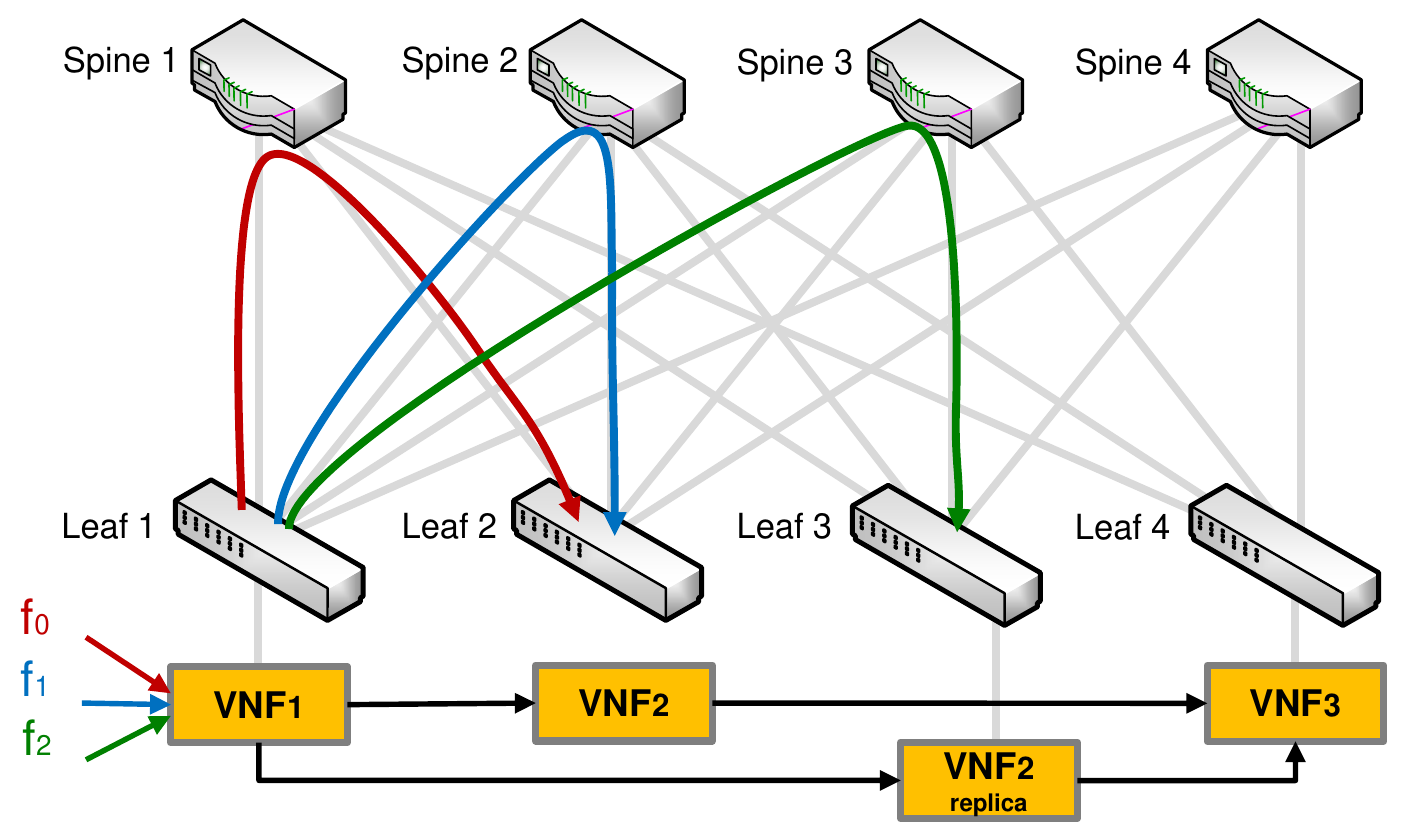}
	\caption{SDN forwarding in leaf-spine fabric}
	\label{leaf-spine}
\end{figure}

\subsection{The SDN scenario}

The same example discussed previously can be also used to show that achieving much better results for network load balancing is possible with SDN in leaf-spine DCN topologies. This is in essence the main reason why the migration towards new DC topologies such as the leaf-spine fabric, in combination with SDN, has been proposed by CORD. A leaf-spine topology example is shown in Fig. \ref{leaf-spine}. Following the same narrative as in the ECMP case, here, each traffic demand is forwarded from Leaf\textsubscript{1} towards the destination node Leaf\textsubscript{4} following the optimal paths. Just like in ECMP scenario, all possible paths are  also here pre-computed, but the optimization method is now able to choose paths to forward the traffic. The traffic engineering can be effectively implemented thanks to the SDN technology used. It should be noted that applying traffic engineering to this model heavily increases the optimization complexity as compared to the ECMP scenario where the randomness of paths is the basic assumption, due to the ECMP protocol-specific features.

\section{Optimization Models and Analysis}

This section formulates two optimization models, ECMP and SDN, subject to their specific constraints, as described next. The notation of all parameters and variables is summarized in Table \ref{parameters}.

\subsection{ECMP forwarding}

The ECMP protocol was designed to load balance a network with multiple paths with the same cost. To achieve that in DCNs, the switches running this protocol use a hash function of the 5-tuple header fields to chose the output port. All packets with the same 5 fields (i.e all packets belonging to the same flow) follow the same path. This randomness on the selection of paths needs to be considered in the optimization model. This is due to the fact that paths are given as input to the optimization model, and are all pre-computed. 

Given that paths are used as input parameters, for each traffic demand, and for all possible server locations, one ECMP path is randomly calculated between two servers. Therefore, the complete set of all possible sets of ECMP paths $\mathbb{P}$ is defined by:
\begin{equation*}
\mathbb{P} = \bigcup_{s = 0}^{|S|-1} \mathbb{P}_s
\end{equation*}
, where $\mathbb{P}_s$ is a set of sets of ECMP paths for service chain $s$. Taking into account that one service chain provides service to multiple traffic demands, and depending on which combination of VNF locations the traffic demand chooses, a set of paths belongs to the set (of sets) $\mathbb{P}_s$:
\begin{multline*}
    \mathbb{P}_s = [ \{ p_\textsubscript{0,0}^s, p_\textsubscript{0,1}^s, ..., p_\textsubscript{0,y}^s\}, \{ p_\textsubscript{1,0}^s, p_\textsubscript{1,1}^s, ..., p_\textsubscript{1,y}^s\},\\
     ..., \{ p_\textsubscript{x,0}^s, p_\textsubscript{x,1}^s, ..., p_\textsubscript{x,y}^s\} ] \quad with \quad \left\{
                \begin{array}{ll}
                  x = [|V_s| \cdot (|X| - 1)] - 1\\
                  y = |\Lambda|-1
                \end{array}
              \right.
\end{multline*}

Here, the parameter $p_\textsubscript{i,j}^s$ is used to represent a path $j$ from the subset of paths $i$. Following the example shown in Fig. \ref{paths} and, for simplicity, only assuming one VNF per TOR, the VNF2 and the replica can be only placed on TOR2, with 2 admissible paths ($p_\textsubscript{0,0}^s$ and $p_\textsubscript{0,1}^s$), or TOR3, with 4 admissible paths ($p_\textsubscript{1,0}^s$, $p_\textsubscript{1,1}^s$, $p_\textsubscript{1,2}^s$ and $p_\textsubscript{1,3}^s$). Then, for every flow, we randomly pre-select one path for each possible allocation of the VNF2. This is, 3 paths per subset $i$ and, therefore, 6 paths in total because there are two subsets. Later, the model will only select one path per flow. So, in general, for one service chain, $\mathbb{P}_s$ has $|V_s| \cdot (|X| - 1)$ subsets of ECMP paths for all possible server locations of VNFs, except when the VNF is placed on the same server, and $|\Lambda|$ paths (one per traffic demand), per subset $i$. 

\subsection{SDN forwarding}

With SDN-based forwarding, all paths between a pair of servers are also pre-computed and the model chooses the optimal path depending on the objective function. In this case the complete set of paths is defined by:
\begin{equation*}
P = \bigcup_{s = 0}^{|S|-1} P_s
\end{equation*}
, where $P_s$ is the set of all possible paths for service chain s. In contrast to ECMP, here there is no random pre-selection of paths, and the model is free to choose the optimal path for any pair of servers.

\begin{table}[!t]
	\renewcommand{\arraystretch}{1.3}
	\caption{Notation}
	\label{parameters}
	\centering
	\footnotesize
	\begin{tabular}{c p{6.5cm}}
		\hline
		\textbf{Parameter} & \textbf{Meaning}\\
		\hline
		$N$ & set of nodes\\
		$X$ & set of servers\\
		$L$ & set of physical links\\
		$Y$ & set of linear cost functions\\
		$S$ & set of service chains\\
		$V$ & set of all VNFs types\\
		$\Lambda$ & set of all traffic demands\\
		$P$ & set of all pre-computed paths\\
		$\mathbb{P}$ & set of all pre-computed ECMP sets of paths\\
		${V_s} \subseteq V$ & ordered set of VNFs in service chain $s$\\
		$\Lambda_s \subseteq \Lambda$ & set of traffic demands of service chain $s$\\
		$P_s \subseteq P$ & set of available paths for service chain $s$\\
		$\mathbb{P}_s \subseteq \mathbb{P}$ & set of ECMP sets of paths of service chain s\\
		$N_p \subseteq N$ & ordered set of nodes traversed by path $p$\\
		$X_n \subseteq X$ & set of servers in node $n$\\
		$T_{p}^\ell$ & 1 if path $p$ traverses link $\ell$\\
		$R_{v}$ & 1 if function $v$ can be replicated\\
		$R_{MAX}$ & maximum number of allowed replicas per service chain\\
		$L_v$ & load ratio of VNF $v$\\
		$E_m$ & penalty ratio due to migration\\   
		$E_r$ & penalty ratio due to replication\\  
		$C_\ell$ & maximum capacity of link $\ell$ \\
		$C_x$ & maximum processing capacity of server $x$ \\
		\hline
                \multicolumn{2}{c}{\textbf{ECMP Model variables}} \\
                \hline
		$r_{i}^s$ & A binary routing variable, 1 if service chain $s$ is using the subset of ECMP paths $i$ \\
		$r_{i \rightarrow j}^{\lambda,s}$ & A binary routing variable, 1 if traffic demand $\lambda$ from service chain $s$ is using its corresponding ECMP path $j$ from the chosen subset $i$\\
		\hline
                \multicolumn{2}{c}{\textbf{SDN Model variables}} \\
                \hline
                $r_{p}^s$ & A binary routing variable, 1 if service chain $s$ is using path $p$ \\
		$r_{p}^{\lambda,s}$ & A binary routing variable, 1 if traffic demand $\lambda$ from service chain $s$ is using the path $p$\\
                \hline
                \multicolumn{2}{c}{\textbf{Common variables}} \\
                \hline
		$f_x^{v,s}$ & A binary variable, 1 if VNF $v$ from service chain $s$ is allocated at server $x$\\
		$f_{x,\lambda}^{v,s}$ & 1 if VNF $v$ from service chain $s$ is being used at server $x$ by traffic demand $\lambda$\\
		$k_\ell$ & utilization cost of link $\ell$\\
		$k_x$ & utilization cost of server $x$\\
		$k_v$ & migration cost of VNF $v$\\
		\hline
		\vspace{0.1cm}
	\end{tabular}
\end{table}
\begin{figure}[!t]
	\centering
	\includegraphics[width=0.4\textwidth]{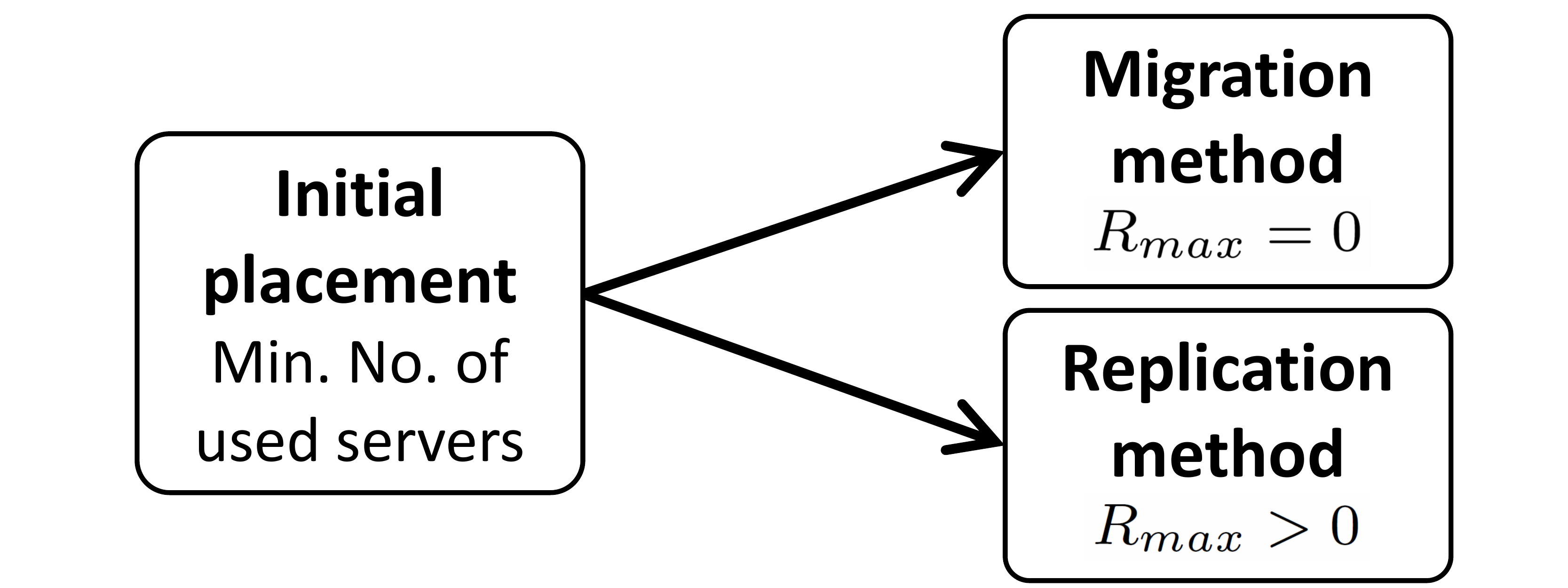}
	\caption{Model execution}
	\label{model}
\end{figure}

\subsection{LP model}

We now present the LP model with the default notation for the SDN model, and the constraints specific for EMCP are notated separately. For both models, two different methods are described, replication and migration. The general objective function, shared by both models and methods, considers the minimization of three different costs: migration, server and link costs, i.e., 
\begin{equation} \label{objective_function}
	minimize: \frac{\alpha}{|V|} \sum_{s \in S} \sum_{v \in V_s} k_v + \frac{1 - \alpha}{|X|} \sum_{x \in X} k_x + \frac{\beta}{|L|} \sum_{\ell \in L} k_\ell
\end{equation}

\par the first two terms, i.e., migration and server cost, are weighted by the $\alpha$ parameter to enable the desired trade-off between the number of migrations and server load balancing. While the link costs are in all cases taken into consideration, the small $\beta$ factor decreases its weight in comparison with the server and migration costs. It should be noted that all three terms are normalized by the total number of functions $|V|$, servers $|X|$ and links $|L|$, respectively. All costs are defined by the resulting values from the piecewise linear cost functions $y_i(u) = a_i \cdot u - b_i$, corresponding to an exponential cost as shown, later, in Fig. \ref{lu_ecmp}. To be noted, that the cost of a replica is implicitly included in the cost of the server where the replica is. This is because the cost to place a replica increases the server utilization due the overhead incurred by the creation of a new VNF. Therefore, the model will only do replication in case the benefit on the server load balancing compensates the cost of creating new VNF.

\subsection{Migration constraints}

Before optimizing the migration, the model generally first optimizes the placement for each source-destination pair of nodes, with a specific ordered set of functions $V_s$ that belong to the service chain ${s \in S}$, as shown in Fig. \ref{model}. In this case, the objective is to minimize the number of servers. In other words, the placement of VNFs determines the initial allocation of VNFs (i.e. $F_{x}^{v,s}$) that will be taken as input parameters for the migration method execution. 

Knowing the initial allocation of VNFs, the migration method specifies the cost of migration in relation to the VNF utilization $u_{v}$ on the server multiplied by a penalty value due the degradation of QoS:
\begin{equation}
	\forall s \in S, \forall v \in V_s, \forall y \in Y: k_v \geq y \Big( u_{v} \cdot QoS_{penalty} \Big)
\end{equation}

The utilization of a VNF (i.e. $u_v$) on a certain server $x$ is determined by the sum of all traffic demands that are using the VNF multiplied by a specific load ratio, which is different depending on the VNF type:
\begin{equation}
	\forall s \in S, \forall v \in V_s, \forall x \in X: u_v = \sum_{\lambda \in \Lambda_s} \frac{ \lambda \cdot F_{x, \lambda}^{v,s} \cdot L_v}{C_x} \leq 1
\end{equation}

The used $F_{x, \lambda}^{v,s}$ term is not a variable, but a parameter specified by the initial allocation of VNFs. In case the migration model determines that a VNF has to be migrated, a penalty value has to be taken into account due to the degradation of the QoS. As an example, if the initial placement specifies that a function $v\textsubscript{1}$ was in server $x\textsubscript{2}$ using the parameter $F_{x\textsubscript{2}}^{v\textsubscript{1}, s\textsubscript{0}}$ = 1, and the migration method specifies that the function is not in this server anymore, by using the variable $f_{x\textsubscript{2}}^{v\textsubscript{1}, s\textsubscript{0}}$ = 0,  then, the QoS penalty value is equal to $E_m$, or 0 otherwise: 
\begin{equation}
	\forall x \in X, \forall s \in S, \forall v \in V_s:
	 QoS_{penalty} = F_{x}^{v,s} (1 - f_{x}^{v,s}) \cdot E_m
\end{equation}
, where $E_m$ is a parameter that can be tuned to determine the penalty of a migration. Finally, since in the migration method replication is not allowed (i.e. $R_{MAX}$ = 0), each service chain has to use exactly one path to forward all traffic demands in the SDN model:
\begin{equation} \label{rmax}
	\forall s \in S: \sum_{p \in P_s} r_p^s = 1
\end{equation}
For the EMCP model the equivalent constraint assures that exactly one subset of ECMP paths $i$ from $\mathbb{P}_s$ is used to forward all traffic demands using the variable $r_{i}^s$, instead.

\subsection{Replication constraints}

In the replication method, the number of active paths for each service chain is related to the possible allocation of VNFs in the servers and, therefore, is constrained by the number of replicas (i.e. $R_{MAX} \geq$ 1). Therefore, 
\begin{equation} \label{rmax}
	\forall s \in S: 1 \leq \sum_{p\in P_s} r_{p}^s \leq R_{MAX} + 1
\end{equation}
Again in the ECMP model, instead of paths $p$, the equivalent constraint considers the variable $r_{i}^s$ where $i$ is a set of ECMP paths from service chain $s$. Then, the constraint limits the number of active sets of paths depending on the number of replicas. 

The next constraint defines how many times a VNF can be replicated, which is determined by the parameter $R_v$: 
\begin{equation}
\forall s \in S, \forall v \in V_s:  \sum_{x \in X} f_x^{v,s} \leq R_v \sum_{p \in P_s} r_{p}^s + 1 - R_v
\end{equation}
If the function can be replicated, then the function can be placed in many servers as active paths the variable $r_{p}^s$ determines. One more time, in the ECMP model the same constraint is considered just changing $P_s$ by $\mathbb{P}_s$ and using the related variable $r_{i}^s$ to determine how many times the VNF can be replicated in relation to how many active $i$ sets of ECMP path are activated. 

\subsection{General constraints}

The rest of the constraints are common to both methods, i.e., replication and migration. Following the objective function (\ref{objective_function}), the server and link costs are respectively defined by:
\begin{equation}
	\forall n \in N, \forall y \in Y: k_x \geq y \big( u_{x} \big)
\end{equation}
\begin{equation}
	\forall \ell \in L, \forall y \in Y: k_{\ell} \geq y \big( u_{\ell} \big)
\end{equation}

The server utilization is calculated adding the utilization of every VNF and the overhead introduced by the creation of the VNF: 
\begin{equation}
     \forall x \in X: u_{x}  = \sum_{s \in S} \sum_{v \in V_s}  u_v + h_v\leq 1
\end{equation}
Then, the calculation of the VNF utilization is defined as:
\begin{equation}
u_v = \sum_{\lambda \in \Lambda_s}  \frac{\lambda \cdot f_{x,\lambda}^{v,s} \cdot L_v}{C_x}
\end{equation}
, where the variable $f_{x,\lambda}^{v,s}$ specifies if a certain traffic demand $\lambda$ is using the VNF $v$ in server $x$. If true, then using the corresponding load ratio $L_v$ for the specific function $v$, the bandwidth from $\lambda$ is added. On the other hand, the overhead is calculated following a linear function:
\begin{equation}
h_v = E_r \cdot u_v + \frac{f_{x}^{v,s}}{C_x \cdot E_r}
\end{equation}
, where the first term adds the percentage of overhead, that increases in relation to the utilization of the VNF, and is pondered by the parameter $E_r$. The second term, also in pondered by $E_r$, adds a fixed percentage which does not depends on the utilization but only on the capacity of the server. This term is the minimum overhead that any VNF has due to its existence. To know if the VNF exist on server $x$, the variable $f_{x}^{v,s}$ is used. In case the VNF is not in the server, both terms are zero. On the other hand, the link utilization is defined as:
\begin{equation}
    \forall \ell \in L: u_{\ell}   = \sum_{s \in  S} \sum_{\lambda \in  \Lambda_s} \sum_{p \in P_s} \frac{\lambda \cdot  r_{p}^{\lambda,s}  \cdot T_{p}^\ell}{C_\ell} \leq 1
\end{equation}
, where the variable $r_{p}^{\lambda,s}$ specifies when a specific traffic demand $\lambda$ is using the path $p$. If true, then the condition $T_{p}^\ell$ checks if path $p$ is traversing the link $\ell$, in order to sum the bandwidth $\lambda$ to the equation. In the equivalent constraint for the ECMP model, the variable $r_{i \rightarrow j}^{\lambda,s}$ returns an specific path $j$ from the subset $i$ used by traffic demand $\lambda$.

The following routing constraint defines that each traffic demand $\lambda$ from each service chain $s$ can only use one path $p$:
\begin{equation}
\forall s \in  S, \forall \lambda \in  \Lambda_s: \sum_{p \in P_s} r_{p}^{\lambda,s} = 1
\end{equation}

In the ECMP model, the variable $r_{i \rightarrow j}^{\lambda,s}$ is also a three-dimensional variable ($s, \lambda, i$) that specifies which set of paths the traffic demand chooses. Then, based on the chosen set, a randomly pre-selected path $j$ is returned for the specific traffic demand and the same previous constraint is applied.

The next constraint assure that when some traffic demand $\lambda$ is using the path $p$, this is activated for the service chain $s$:
\begin{equation} \label{routing_1}
\forall s \in  S, \forall \lambda \in  \Lambda_s, \forall p \in P_s : r_{p}^{\lambda, s} \leq r_{p}^{s} \leq \sum_{\lambda' \in \Lambda_s} r_{p}^{\lambda', s}
\end{equation}
One more time the iteration $\forall p \in P_s$ is $\forall p_i^s \in \mathbb{P}_s$ for the ECMP model, and the variables $r_{i \rightarrow j}^{\lambda,s}$ and $r_{i}^{s}$ are used instead.

Then, the next constraint allocates all VNFs from the service chain $s$ in the activated path using the variable $f_{x,\lambda}^{v,s}$:
\begin{equation}
\forall s \in  S, \forall p \in P_s, \forall \lambda \in \Lambda_s, \forall v \in  V_s: r_{p}^{\lambda, s} \leq \sum_{n \in N_p} f_{x,\lambda}^{v,s} 
\end{equation}
, where $N_p$ is an ordered set of servers traversed by path $p$. Note that for the ECMP model, the variable $r_{i\rightarrow j}^{\lambda, s}$ specifies the path $j$ when a traffic demand $\lambda$ chooses the subset $i$. 

The rest of constraints assure the proper activation of VNFs in the correct order for each traffic demand. First, the next constraint specifies that each traffic demand $\lambda$ has to traverse an specific function $v$ in only one server:
\begin{equation}
\forall s \in S, \forall v \in  V_s, \forall \lambda \in \Lambda_s: \sum_{x \in  X} f_{x,\lambda}^{v,s} = 1
\end{equation}
Then, similarly to (\ref{routing_1}), the next constraint allocates the function $v$ on server $x$ when at least one traffic demand is using it:
\begin{equation}
\forall s \in  S, \forall v \in  V_s, \forall x \in X, \forall \lambda \in \Lambda_s: f_{x,\lambda}^{v,s} \leq f_x^{v,s} \leq \sum_{\lambda' \in  \Lambda_s} f_{x,\lambda'}^{v,s} 
\end{equation}
\par Finally, since each service chain is composed by a certain ordered set of VNFs, each traffic demand has to traverse them in the correct order, i.e.,
\begin{multline}
	\forall s \in S, \forall \lambda \in \Lambda_s, \forall p \in P_s, \forall v \in {V_s}, \forall n \in N_p, \forall x \in X_n: \\
	\Bigg( \sum_{m = 0}^{n} \sum_{y \in X_m} f_{y, \lambda}^{(v-1),s} \Bigg) - f_{x, \lambda}^{v,s} \geq r_{p}^{\lambda,s}  - 1 \quad if \quad v>0
\end{multline}

, where for every traffic demand $\lambda$, the function $v$ is allocated at server $x$, only if the previous function $v-1$ of the same service chain $s$ is already allocated in any of the previous available servers $y$ from the activated path $p$. Note that $N_p$ is an ordered set of nodes traversed by the path $p$ and $X_n$ is the set of servers running on node $n$. Finally, the same constraint is applied in the ECMP model, just iterating over $\mathbb{P}_s$ and considering the variable $r_{i\rightarrow j}^{\lambda, s}$, instead.

\section{Performance Evaluation}

This section shows the results from the LP models implemented with Gurobi Optimizer \cite{gurobi}. To test the ECMP model, we analyze a fat-tree topology with 4 pods and 4 servers per TOR (32 servers in total). For the SDN model, we use a leaf-spine topology with 4 leaf switches and 4 servers per leaf (16 servers in total). In all cases, we assume a server capacity of 1000 units and 1 Tbps of link capacity. For both scenarios, we assume that every source-destination pair of servers instantiate its own SFC and all nodes are able to allocate an unlimited number of VNFs, and are only restricted by the resource capacity of the server. In the SDN model,  the paths are optimized, which increases the computing complexity as compared to the ECMP model where paths are pre-computed. For that reason, we use a smaller topology to reduce computing time. Also for simplicity, we assume that servers are in TOR or in leaf switches, and we do no the interconnects between TOR switches and servers. The simulations show the results after the initial placement of VNFs, where the model attempts to migrate or replicate VNFs to improve the server and link utilization.
 
\subsection{The ECMP scenario}

In the fat-tree topology, each source-destination pair of servers randomly generates between 10 and 20 connections, each connection within the interval [1, 30] Gbps. The SFC chosen to test is composed by 3 VNFs, where only the second VNF in the chain can be replicated. The load ratio (i.e. $L_v$), which indicates the required computing resources to perform the task depending on the processed traffic, is 0.2 for the first and last VNF, and chosen 1 for the second VNF. The reason to choose this configuration is to analyse a common case where one VNF is compute intensive, while the others just act as a load balancers, with a comparably low compute intensive task.

\begin{figure}
	\centering
	\includegraphics[width=0.5\textwidth]{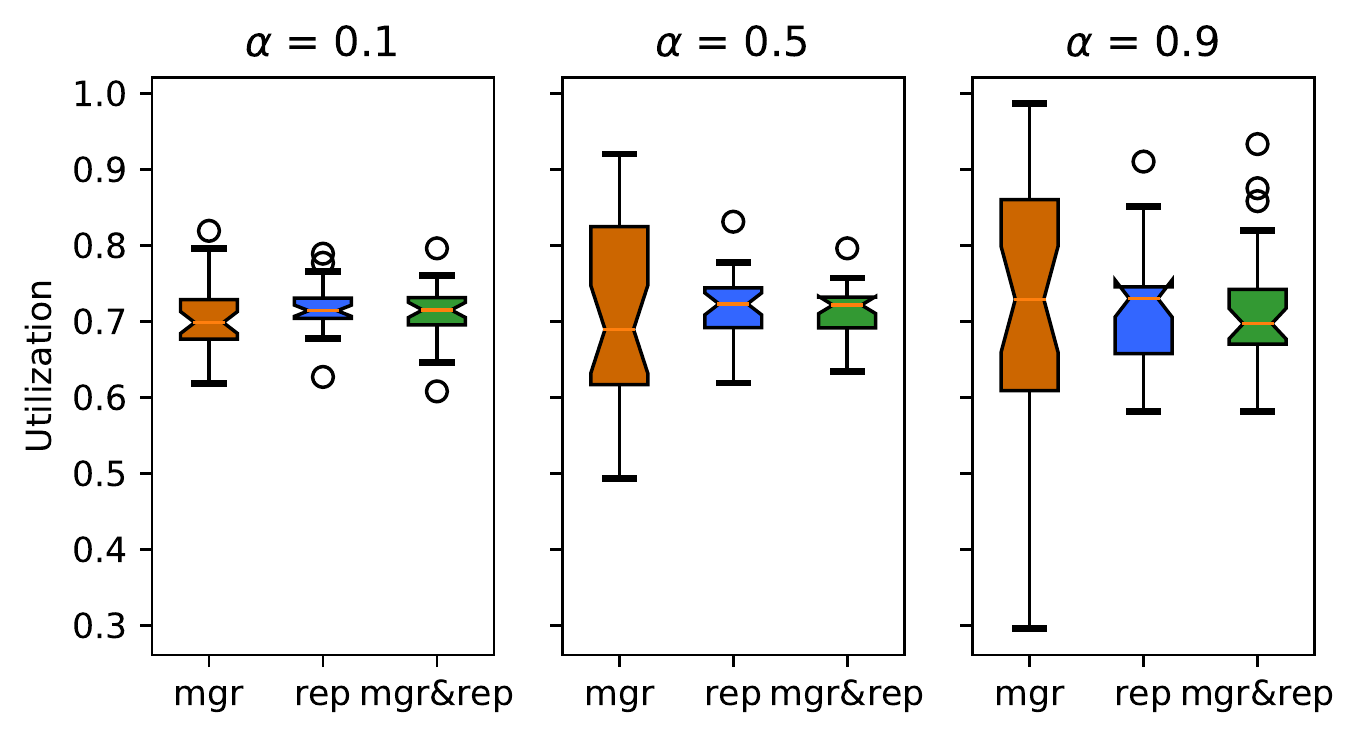}
	\caption{Fat-tree topology - Server Utilization}
	\label{su_ecmp}
\end{figure}

Fig. \ref{su_ecmp} shows the server utilization for cases where only migrations are used ($mgr$), only replications are used ($rep$) and when both migrations and replication can be used ($mgr\&rep$). When $\alpha$ = 0.1, which means that the server load balancing has more weight in the objective function (see equation (\ref{objective_function})), in the $rep$ case, servers are better load balanced than in the $mgr$ case, however the average server utilization increases due to 17 replicas (see Table \ref{number}). The $mgr\&rep$ case, which requires 31 migrations and 9 replicas, slightly outperforms both previous cases. When $\alpha$ = 0.9 and we only allow replicas, the model clearly outperforms the $mgr$ case, which is quite obvious because we are trying to minimize the migrations. The interesting case is when $\alpha = 0.5$ and the $mgr\&rep$ case improves by using 26 migrations and 13 replicas. Here, it is clear how the balancing between the number of migrations and replications benefits the server load balancing, even when the mean values increase due to the overhead introduced by the replicas. On the other hand, only performing migrations achieves an acceptable server load balancing when the number of migrations is high, while only doing replication performs well in all cases but increases the average server utilization, because of the replication overhead.

\begin{table}[]
\centering
\caption{Number of migration versus number of replications}
\label{number}
\begin{tabular}{c c|c|c|c}
\multicolumn{1}{c}{}         &              & $\alpha$ = 0.1 & $\alpha$ = 0.5 & $\alpha$ = 0.9\\ \hline
\multirow{3}{*}{\textbf{ECMP}} & mgr &   41     &  34 &  31 \\ \cline{2-5} 
                               & rep &    17    &  17 & 15  \\ \cline{2-5}
                               & mgr\&rep & 31-9  & 26-13  &  29-9 \\ \hline\hline
\multirow{3}{*}{\textbf{SDN}}  & mgr &    10    &  7  &   7\\ \cline{2-5} 
                                & rep &   11     & 11   &  6 \\ \cline{2-5} 
                               & mgr\&rep & 11-12  &  9-12  &  8-17 \\ \hline
\end{tabular}
\end{table}

\begin{figure}
	\centering
	\includegraphics[width=0.49\textwidth]{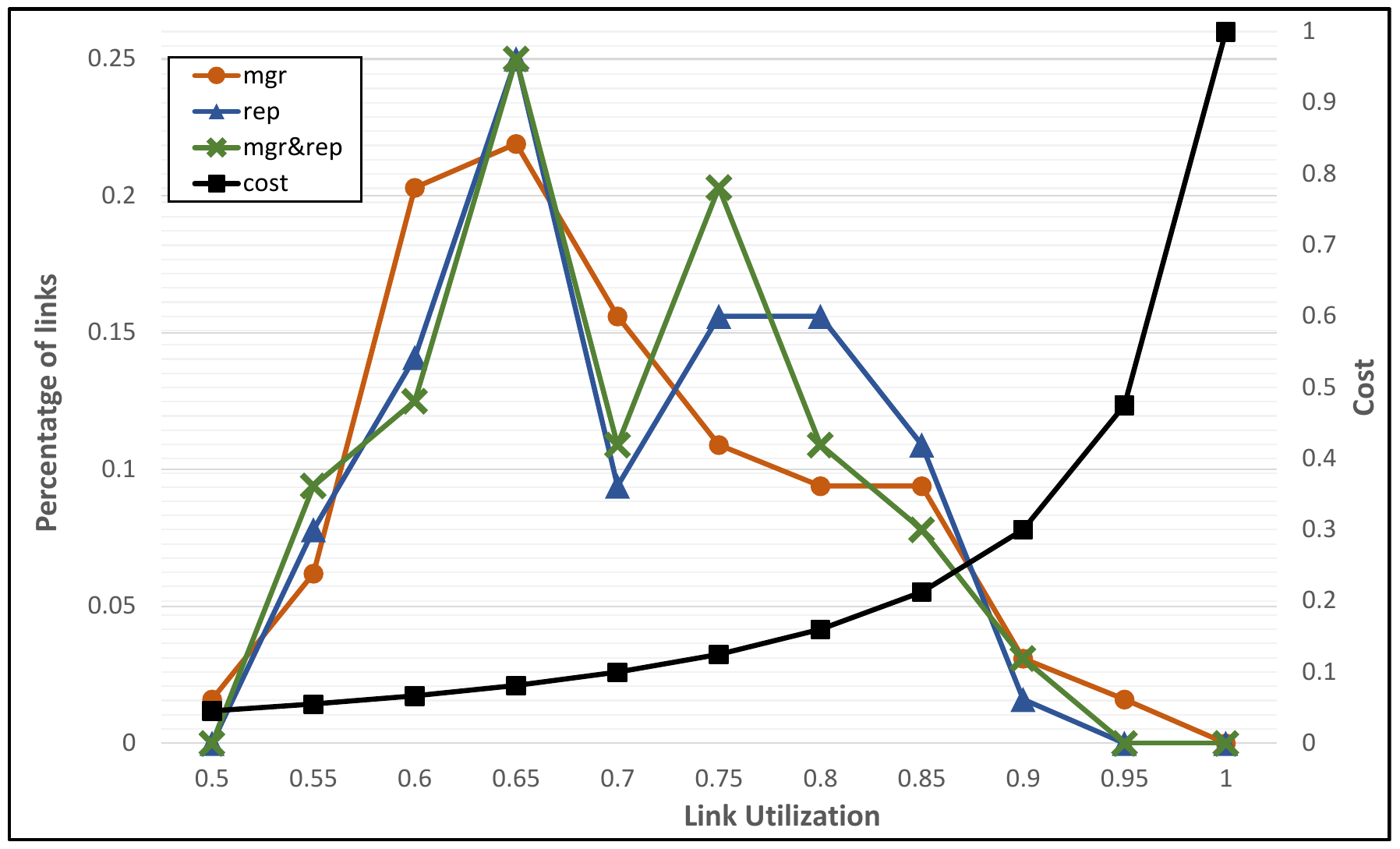}
	\caption{Fat-tree topology - Link Utilization}
	\label{lu_ecmp}
\end{figure}

Since the network load balancing is a relevant aspect in DCNs, the minimization of all link costs is always included either in the migration or the replication method, but with less weight, as shown in the previous section. Fig. \ref{lu_ecmp} shows the link utilization in relation to the percentage of links in the fat-tree topology using EMCP protocol for $\alpha = 0.5$. Here, the objective function tries to decrease the number of overload links applying an incremental cost which emulates an exponential function, as also shown in Fig. \ref{lu_ecmp}. The reason to only show one value of $\alpha$ is because the results are quite similar for all $\alpha$ values. This is because the ECMP protocol is not flexible enough to optimize the link load balancing as later shown for the SDN case, where the paths are optimized also for network load balancing.

As shown in Fig. \ref{lu_ecmp}, all three cases perform relatively similar having the $mgr$ case some overloaded links with 95\% of utilization and the $rep$ case working better when the utilization is over 90\%. In general, the $mgr\&rep$ case performs better than the other two cases by reducing the total number of links over 80\% of utilization. The reason for this is that replicas are increasing the number of admissible paths to forward traffic and therefore decreasing the probabilities to create bottlenecks in the network. In this case, however, the use of replica or migrations do not have an obvious benefit on the network load balancing due to the limitation intrinsic to the ECMP protocol.

\subsection{SDN scenario}

In the SDN scenario, the traffic demands are generated at the leaf switch 1 towards the leaf switch 4, and vice versa. In this case, the leaf switch 1 and 4 allocate 3 service chains each, with a number of traffic demands per service chain between 6 and 12 demands, with each demand within the interval [70-110] Gbps. The chosen SFC to test the topology is also here composed by 3 VNFs, where only the second one can be replicated. The load ratios are similar to the previous case, with 0.4, 0.9 and 0.4 values, respectively. 

\begin{figure}
	\centering
	\includegraphics[width=0.5\textwidth]{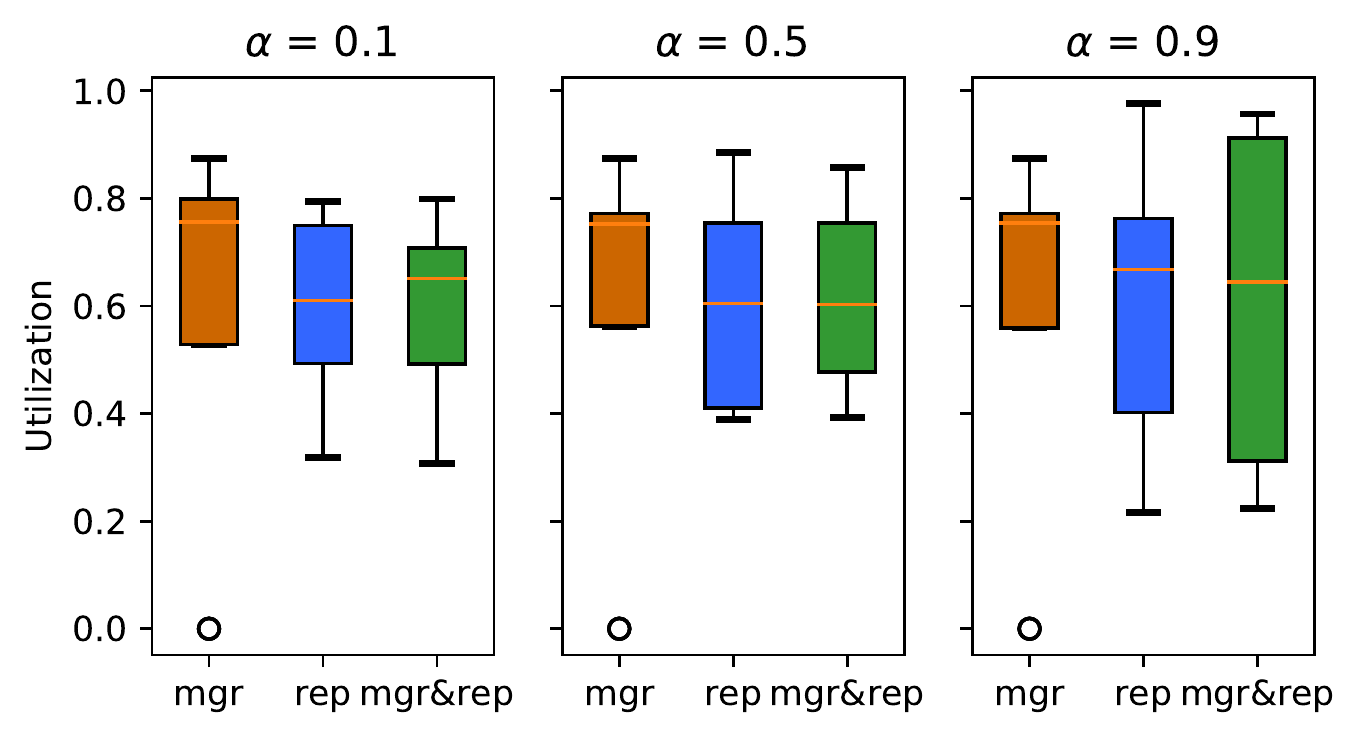}
	\caption{Leaf-Spine topology - Server Utilization}
	\label{su_sdn}
\end{figure}

The server utilization for the three $\alpha$ values is shown in Fig. \ref{su_sdn} and the number of migrations and replications in Table \ref{number}. In the case of migrations only ($mgr$) and to load balance the server utilization, the results are quite similar for all alpha values. In case of replications only ($rep$), the results are promising, for $\alpha=0.1$ and $\alpha=0.5$,  both decreasing the mean utilization by using 11 replicas, as compared to the case of migrations only. In the $mgr\&rep$ case, where both methods are allowed, for $\alpha=0.1$ and by using 11 migrations and 12 replications, the results are slightly worse than in the $rep$ case, and for $\alpha=0.5$ and by using 9 migrations and 12 replications, slightly better than in the $rep$ case. When alpha is $0.9$ and the model performs 6 replicas in the $rep$ case and, 8 migrations and 17 replications in the  $mgr\&rep$, both cases decrease the average server utilization, but increase the number of overloaded servers. 

By analyzing the results in this case study, we can see that performing migrations only negatively affects the server utilization as compared to the case where only replications are allowed. Combining both replications and migrations have better results when the objective is balanced by using the $\alpha$ factor. The practical reason to explain this is due to the topology itself and where the traffic demands are originated. In leaf-spine topologies like it was proposed by OpenCord \cite{cord}, the demands are only originated in the extremes of the topology, which releases resources in the middle allowing the model to have more options where to allocate replicas on underused servers. 

\begin{figure}
	\centering
	\includegraphics[width=0.5\textwidth]{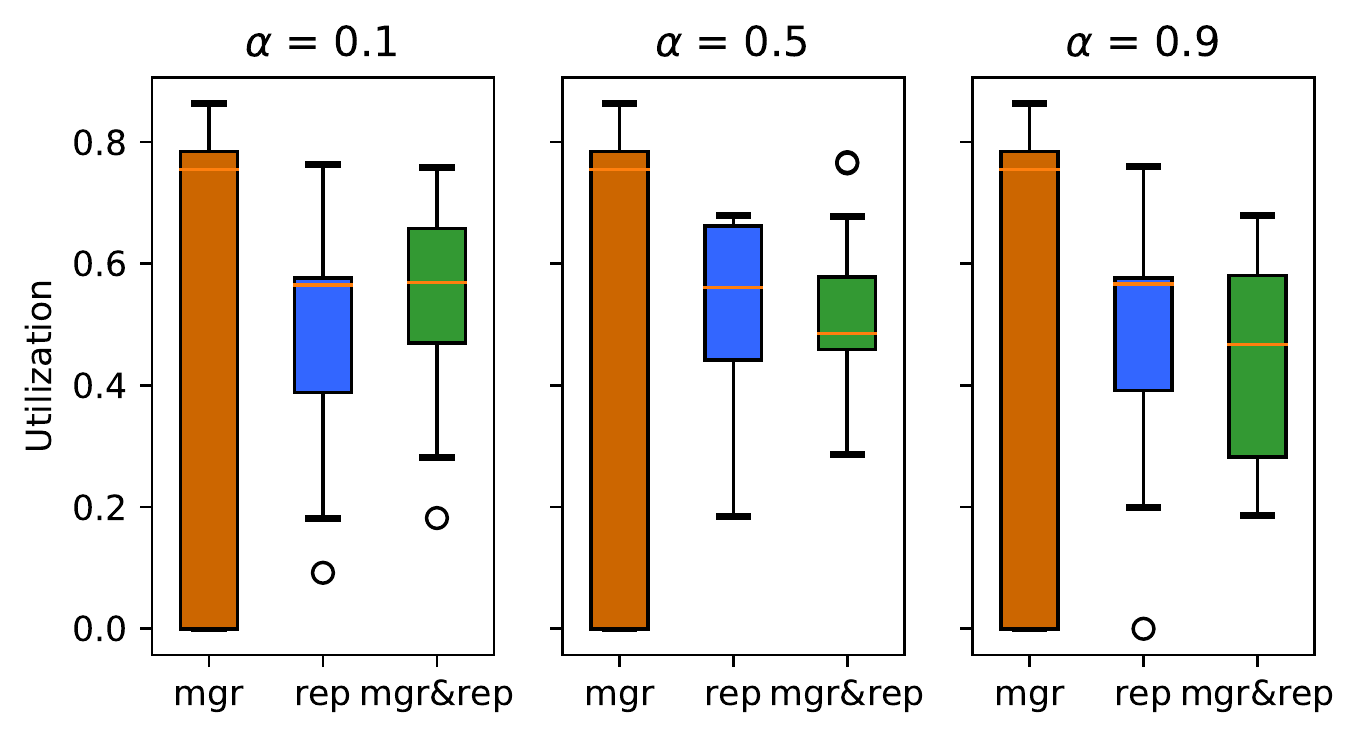}
	\caption{Leaf-Spine topology - Link Utilization}
	\label{lu_sdn}
\end{figure}

The link utilization is shown in Fig. \ref{lu_sdn}. As expected, the network load balancing for the $mgr$ case is rather poor for all $\alpha$ values. This behavior can be explained so that, by performing a migration, we are not addressing the network bottlenecks. At the same time, in $rep$ case, the network is comparably better balanced by decreasing the average link utilization due to the alternative  paths available due to the replicas. The $mgr\&rep$ case performs worse than the $rep$ case when $\alpha = 0.1$, and considerably better when $\alpha = 0.5$ and $\alpha =0.9$. This is not straightforward to explain, and requires further study, but it confirms that the combination of migrations and replications has benefits also for network load balancing.

\section{Conclusions}
To balance the server allocation strategies and QoS, this paper proposed the usage \emph{replications} of VNFs to reduce migrations in DCNs. We proposed a LP model to study a trade-off between replications, which while beneficial to QoS require additional server resources, and migrations, which while beneficial to server load management can adversely impact the QoS. The results show that, for a given objective, the replications can reduce the number of migrations and can also enable a better server and data center network load balancing. For future work, we plan to extend the model to study how migrations affect the service chains when multiple VNFs are contained in a single VM and what role replications play here. The selection of SFCs matching to a more realistic scenarios is also taken into account as future work.

\section*{Acknowledgment}
This work has been performed in the framework of SENDATE-PLANETS (Project ID C2015/3-1), and it is partly funded by the German BMBF (Project ID 16KIS0470).

\bibliographystyle{IEEEtran}
\bibliography{mylib}

\end{document}